# Compositional dependence of magnetic damping in sputter-deposited $Co_xFe_{1-x}$ thin films

**Samanvaya S. Gaur[a], Rosa Diaz[b], Ernesto E. Marinero[a]\*,**

[a]School of Materials Engineering, Purdue University, West Lafayette, IN, USA

[b]School of Electrical and Computer Engineering, Purdue University, West Lafayette, IN, USA

## Abstract:

$Co_{25}Fe_{75}$ ferromagnetic films exhibit ultralow magnetic damping. The magnetic damping dependence of $Co_{1-x}Fe_x$ thin films over a Co composition (23 – 36 %) is here reported. The thin film structures were sputter deposited at ambient temperature and FMR measurements in both in-plane and out-of-plane geometries were utilized to measure magnetic damping parameters, $\alpha_{tot}$, which include intrinsic damping and contributions from spin pumping. The damping parameters, $\alpha_{tot}$, decrease as the Co content is increased, except for Co = 31 %. The smallest values of $\alpha_{tot}$ correspond to alloys exhibiting interface perpendicular magnetic anisotropy. A value of $\sim 0.91 \times 10^{-3}$ was measured for $Co_{36}Fe_{64}$, whereas $\alpha_{tot}$ for $Co_{31}Fe_{69}$ was measured as $\sim 2.04 \times 10^{-3}$, this composition exhibits the largest in-plane anisotropy.

HAADF-STEM cross-section analysis of the $Co_{36}Fe_{64}$ thin film stack revealed Cu interdiffusion into the magnetic layer. The degree of interdiffusion was found to be up to 7x higher at grain boundaries as compared the bulk of the polycrystalline grains. The incorporation of Cu into the ferromagnetic layer adversely impacts magnetic damping. Reducing impurities in the magnetic layer by improving the growth chamber base pressure resulted in a reduction of magnetic damping of 18 %. The diffraction analysis revealed that the primary growth direction of $Co_{36}Fe_{64}$ is (101) and that of Cu buffer layer is (111), these planes are perpendicular to their respective ($\bar{1}$01) planes and for this composition the lattice mismatch was determined to be 0.9325 %. The lattice mismatch decreases with increasing Co content and hence the lattice strain. The diffusion of Cu into the ferromagnet creates magnon scattering centers and local changes in magnetic properties. Both factors negatively influence magnetic damping.

This work is suggestive of potential avenues to further reduce magnetic damping in Co-Fe alloy thin films by controlling the alloy composition, impurities, strain relaxation and interdiffusion from seed layers which are required for crystallographic control.

## Introduction

With an ever-increasing demand for energy efficient information processing, non-conventional computing platforms based on spin wave propagation have emerged as a potential solution. Current electronic devices are hindered by ohmic losses resulting in excessive heating requiring device cooling. This is worsened as the density of integrated devices increases. Magnonics, on the other hand, which involves magnons as information carriers (analogous to electrons) but without ohmic losses, provides an energy efficient method for computation. Furthermore, the wave-nature of magnons, opens opportunities for information encoding and processing that exploit the amplitude, phase and coherence of magnons. A critical requirement for magnonic materials is to exhibit low magnetic damping constants to enable magnon long-range transport in magnetic waveguides.[1,2] Ferrimagnetic insulators such as yttrium iron garnets (YIG), are employed as the magnonic material of choice on account of their ultralow magnetic damping constant, tunability of magnetic properties and low microwave losses.[3-5] However, liquid phase epitaxial growth on single crystal substrates of gadolinium gallium garnets is required at deposition/annealing temperatures > 700 °C to attain the lowest damping factors.[2,6] Therefore, presently these materials and substrates are incompatible with CMOS-based fabrication, hindering the implementation of magnonic devices as computational hardware.

\*Corresponding author, eemarinero@purdue.edu

Recently, the ferromagnetic metal alloy, $Co_{25}Fe_{75}$, has been identified as a potential candidate for nanoscale magnonic applications as it exhibits magnetic damping parameters as low as 0.002 and it can be grown on Si wafers employing sputter deposition.[1,7-9] Magnon propagation lengths up to 20 μm have been reported.[9] In addition, epitaxial growth of $Co_{25}Fe_{75}$ on MgO and $MgAl_2O_4$ substrates has resulted in magnetic damping parameters of 0.00071 and 0.001 respectively.[10]

Inspired by such reports, in this work we report measurements of magnetic damping parameters in $Co_{1-x}Fe_x$ films over a narrower composition range than that reported in reference 7. We investigate the compositional dependence of magnetic damping parameter on interface magnetic anisotropy as well as the effect of impurities. Cross section HAADF-STEM is employed to identify the growth orientation of both the ferromagnetic and seed layers and determine the degree of lattice mismatch (lattice strain) between their respective crystallographic planes parallel to substrate plane.

The magnetic damping parameters, $\alpha_{tot}$, extracted from ferromagnetic resonance (FMR) measurements conducted in this work include intrinsic and spin pumping contributions, and are not affected by radiative damping interference.

**Experimental**

Binary alloy thin films were deposited by DC magnetron co-sputtering from pure Co and Fe targets in a chamber with a base pressure < $9 \times 10^{-8}$ Torr. The ferromagnetic thin films were grown on, and capped with, Ta/Cu bilayers on oxidized Si wafers. The layered structures fabricated were: Si[100](500 μm)/SiO₂(90 nm)/Ta(3 nm)/Cu(3 nm)/Co₁₋ₓFeₓ(10 nm)/Cu(3 nm)/Ta(3 nm). The sputter pressure used for all layers was 5 mTorr in all depositions. Profilometry was employed to calibrate target deposition rates and thin film thicknesses. The target powers employed were 5 W to 8 W and 30 W to 34 W, respectively for Co and Fe. The Ta and Cu thin films were deposited in-situ with a power of 5 W. The target-substrate spacing was 11 cm and the angle of the targets from the normal to the substrates was ≈ 13 degrees. The substrate holder was rotated at 5 rpm to ensure thin film uniformity.

X-ray Fluorescence (Malvern Panalytical Epsilon 4) was employed to determine the magnetic thin film compositions. Magnetic properties, saturation magnetization, coercivity and hard-axis fields, were measured using a SQUID magnetometer (Quantum Design MPMS-3 SQUID Magnetometer). Magnetic damping constants were derived using a broadband ferromagnetic resonance spectrometer (Quantum Design DynaCool PPMS vacuum chamber and NanOsc Instruments CryoFMR spectrometer). Both in-plane (ip) and out-of-plane (oop) measurements were performed and the samples were placed face down on the co-planar waveguide. Kapton tape

*Corresponding author, eemarinero@purdue.edu

(~ 0.07 mm thick) was placed between the sample surface and the waveguide to avoid radiative damping.[7] Out-of-plane geometry (oop) measurements, eliminate two magnon scattering contributions to magnetic damping. In this work spin pumping contributions to damping are not subtracted. A Themis Z Aberration Corrected Scanning/Transmission Electron Microscope was used for atomic level composition and diffraction analysis.

**Results and Discussion**

Ferromagnetic resonance FMR measurements are conducted at a range of fixed frequencies while sweeping the external magnetic field ($H_{DC}$). A representation of the spin dynamics and of the orthogonal DC and RF fields are shown in Figure 1 (a), a constant $\mathbf{H_{DC}}$ is applied to the in-plane magnetized sample of magnetization, $\mathbf{M}$. The coplanar waveguide produces a small perturbing field, $\mathbf{H_{RF}}$, applied perpendicular to the external field, $\mathbf{H_{DC}}$, which induces magnetization precession around the equilibrium position at resonance condition (determined from the Kittel equation). Due to the magnetic damping, the precessing magnetization eventually spirals around towards the equilibrium position. The magnitude of the magnetic damping parameter determines how quickly the magnetic material responds to external perturbations.

In FMR studies with the DC field applied parallel to the plane of the film (in-plane configuration), the resonance field, $\mathbf{H_{res}}$, is given by the Kittel equation[8]:

$$f = \frac{\gamma \mu_0}{2\pi} \sqrt{[M_{eff} + H_{res} + H_k][H_{res} + H_k]} \qquad \text{(Equation 1)}$$

where $f$ is the Kittel frequency (in GHz), $\gamma$ is the material's gyromagnetic ratio, $\gamma = \frac{g \mu_B}{\hbar}$, $\mu_0$ is the vacuum permeability, $\mu_B$ is the Bohr magneton and $\hbar$ is the reduced Planck constant. $\mathbf{M_{eff}}$, $\mathbf{H_{res}}$ and $\mathbf{H}_k$ are the effective magnetization ($M_{eff} = M - H_k^{\perp}$, where $M$ is the saturation magnetization and $H_k^{\perp}$, is the perpendicular anisotropy field), the resonance field, and the uniaxial in-plane magnetic anisotropy field ($H_u^{ip}$), respectively. The value of the Landé factor, $g$, utilized in all FMR measurements is the experimental value reported for $Co_{1-x}Fe_x$ alloys[11] over a wider composition range with broad band FMR over a frequency range of 10 to 40 GHz. As the FMR frequency limit of our setup is 18 GHz, we opted to use published measured values for the composition range investigated in this work. The calculated value of the ratio $\frac{\gamma}{2\pi}$ is ~ 29.5 GHz/T and this value was employed in all FMR (ip) and (oop) measurements.

A schematic illustration of ip-geometry FMR measurement is shown in Figure 1 (a). The purple dots represent the presence of potential defects in the polycrystalline thin films, such as impurities, crystalline defects, grain boundaries, magnetic anisotropy and moment fluctuations. The magnetization dynamics and magnetic damping can be expected to be influenced by the presence of such defects.

*Corresponding author, eemarinero@purdue.edu

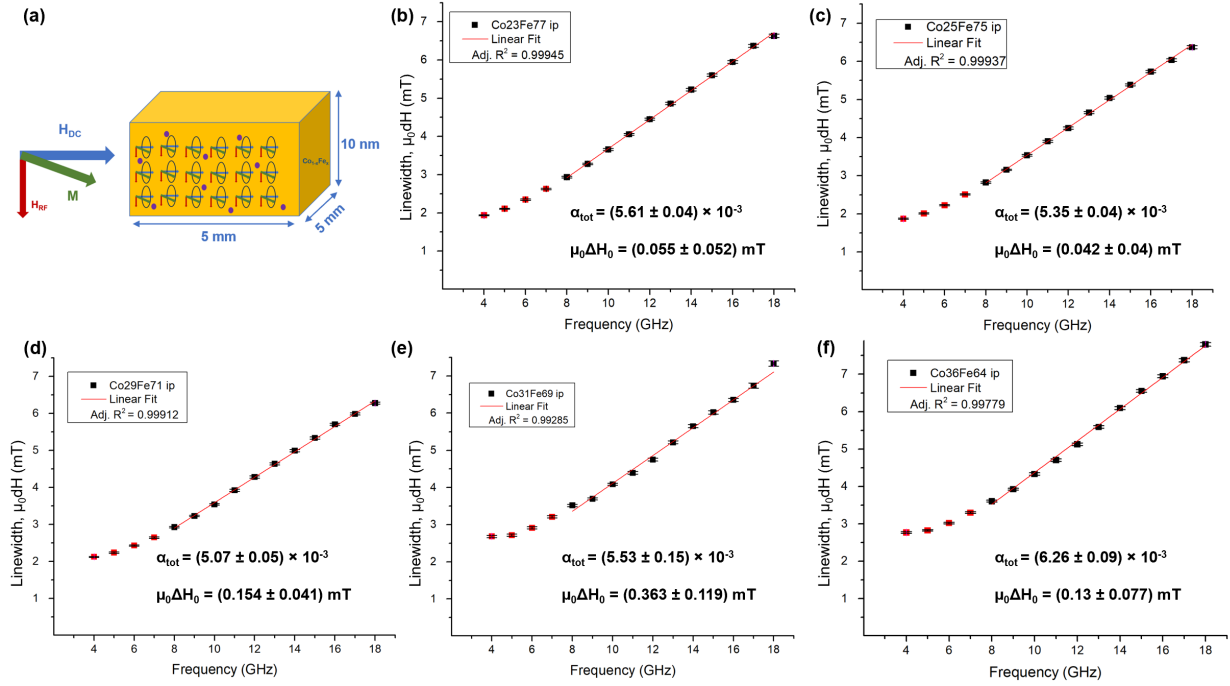

**Figure 1.** In plane linewidth vs. frequency: (a) Representation of in-plane magnetization dynamics in the thin film $Co_{1-x}Fe_x$ samples. The purple dots represent defects in the film that act as sites resulting in two magnon scattering contributions to magnetic damping, (b) $Co_{23}Fe_{77}$, (c) $Co_{25}Fe_{75}$, (d) $Co_{29}Fe_{71}$, (e) $Co_{31}Fe_{69}$ and (f) $Co_{36}Fe_{64}$. The measurement error bars are longer than the size of the data symbols (see above) and originate from fitting the raw data to Equation 2. The red points were excluded from the determination of magnetic damping parameters on account of two magnon scattering at frequencies lower than 8 GHz (see main text).

To determine the magnetic damping parameter ($\alpha_{tot}$), a sequence of measurements of the microwave absorption and its derivative are performed at different frequencies of the microwave field, $H_{RF}$. The results are fitted to the following equation to extract $H_{res}$ and the linewidth (dH)[12]:

$$\frac{dP}{dH} = K1 \frac{4dH(H-Hres)}{[4(H-Hres)^2+(dH)^2]^2} - K2 \frac{(dH)^2-4(H-Hres)^2}{[4(H-Hres)^2+(dH)^2]^2} + \text{Slope H} + \text{Offset} \quad \text{(Equation 2)}$$

where H is the applied external field, whose direction is perpendicular to the microwave field and its magnitude is varied to sweep through the resonance condition, dP/dH is the derivative of the absorbed microwave power at the FMR condition, dH and $H_{res}$ are parameters derived from fitting the raw data to Equation 2. Also, K1, K2, Slope H and Offset are other constants derived from the fit. To determine magnetic damping parameters, dH is plotted vs. RF frequency as shown in Figure 1. The magnetic damping parameter $\alpha_{tot}$, is determined from the slope of the plot[8]:

$$dH = \Delta H_0 + \frac{4\pi f \alpha}{\gamma} \quad \text{(Equation 3)}$$

where, $\Delta H_0$ is the frequency-independent term that corresponds to inhomogeneous broadening due to non-uniformities present in the film.

As observed in Figure 1, the linewidth (dH) vs. frequency plots are linear except at frequencies < 8 GHz, this is attributed to inhomogeneous broadening due to two magnon scattering. The broadening results from the inability of the external magnetic field to fully align the magnetization along the applied field direction.[10] It is evident from the magnetic hysteresis plots provided in the *supplementary information,* Figure S1, that it takes in excess of ~ 0.04 T (external field) to completely align the magnetization of $Co_{1-x}Fe_x$ alloys in the ip geometry along the direction of the applied field. Further, from Figure S4 and Table S1 of *supplementary information*, it is evident that the resonance fields below 8 GHz are smaller than 0.04 T (external field), which means that the range of external fields applied are smaller than 0.04 T for low frequencies (4 to 7 GHz), leading to non-linearity in linewidth vs frequency plots in Figure 1. Thus, in our measurements only the results for frequencies from 8 GHz to 18 GHz were employed for determination of magnetic damping parameters.

*Corresponding author, eemarinero@purdue.edu

The data shown in Figure 1 is employed to derive magnetic damping parameters (from the slope of the linewidth vs. frequency) and inhomogeneous broadening, $\mu_0 \Delta H_0$ (from the intercept with the y-axis of the linear fit of linewidth vs. frequency).

The results from Figure 1 indicate that the measured $\alpha_{tot}$ for $Co_{25}Fe_{75}$, $(5.35 \pm 0.04) \times 10^{-3}$, is slightly lower than the previously reported value of $(5.6 \pm 0.2) \times 10^{-3}$ in ref. 8 and that the lowest magnetic damping parameter measured in the ip configuration, $(5.07 \pm 0.05) \times 10^{-3}$ corresponds to $Co_{29}Fe_{71}$.

For FMR measurements in the out-of-plane (oop) configuration, the following equation is employed to estimate $H_{res}$[8]:

$$f = \frac{\gamma \mu_0}{2\pi} \left[ H_{res} - M_{eff} \right] \qquad \text{(Equation 4)}$$

As shown in Figure S1, magnetic fields exceeding 2.5 T should be sufficient to align the magnetization in the hard axis direction. However, misalignment between the sample and the external field direction in FMR measurements, resulting from sample positioning on the FMR waveguide can influence the external field required. Thus, we conducted coarse linewidth measurements for the values of the external fields of 3 T, 3.5 T and 4 T.

The measurement sequence in the FMR measurements in the oop geometry was the following, after the frequency stabilization, a 4 T field along the hard axis was applied, next, the field was swept from high to low in 0.5 mT steps. The range of the magnetic field range were set in accordance with the value of $H_{res}$ predicted by the Kittel equation. Experimental values of $H_{res}$ can differ from those predicted by the Kittel equation on account of misalignment of the sample while placing it on the coplanar waveguide. To circumvent this, a quick field scan is first conducted to determine $H_{res}$, then the field sweep range (**$H_{DC}$**) is chosen to span the entire FMR response for each frequency (see *supplementary information* for its extraction). FMR measurements in the oop configuration are shown in Figure 2.

Non-linearities are evident in dH vs. frequency plots due to incomplete alignment of the magnetization with the external field applied in the hard axis direction. This is clearly observed in in Figure 2 (c) for measurements at < 9 GHz in the $Co_{25}Fe_{75}$ sample. Below 6 GHz, no FMR signals were observed as the external magnetic field applied is insufficient to overcome the demagnetizing energy. For $Co_{25}Fe_{75}$ alloy ($M_s = 2.215$ T), the applied field at 6 GHz was swept from 2.12 T to 2.075 T and at 18 GHz it ranged from 2.615 T to 2.573 T. The entire range of applied fields employed in this study are provided in the *supplementary information* Table S1. The start and stop magnetic fields applied < 14 GHz are smaller than the field required to fully align them in the hard axis direction (see *supplementary information* Table S2 for details on the field range employed for $Co_{1-x}Fe_x$ samples at 14 GHz). Incomplete alignment of the magnetization in the hard axis direction results in larger linewidth values at frequencies < 14 GHz and lead to the non-linear behavior below 14 GHz and were not employed in oop FMR for other compositions. The highest frequency, 18 GHz, of the measurements is the frequency limit of the FMR equipment employed. Nevertheless, measurements by Schoen et al.[7] on identical thin film stacks comprising $Co_{25}Fe_{75}$ and $Co_{20}Fe_{80}$ over a wider frequency range that includes the $14 – 18$ GHz range of our experiments, show a linear behavior of linewidth vs. frequency, and the linewidths values reported are comparable to our measured values.

*Corresponding author, eemarinero@purdue.edu

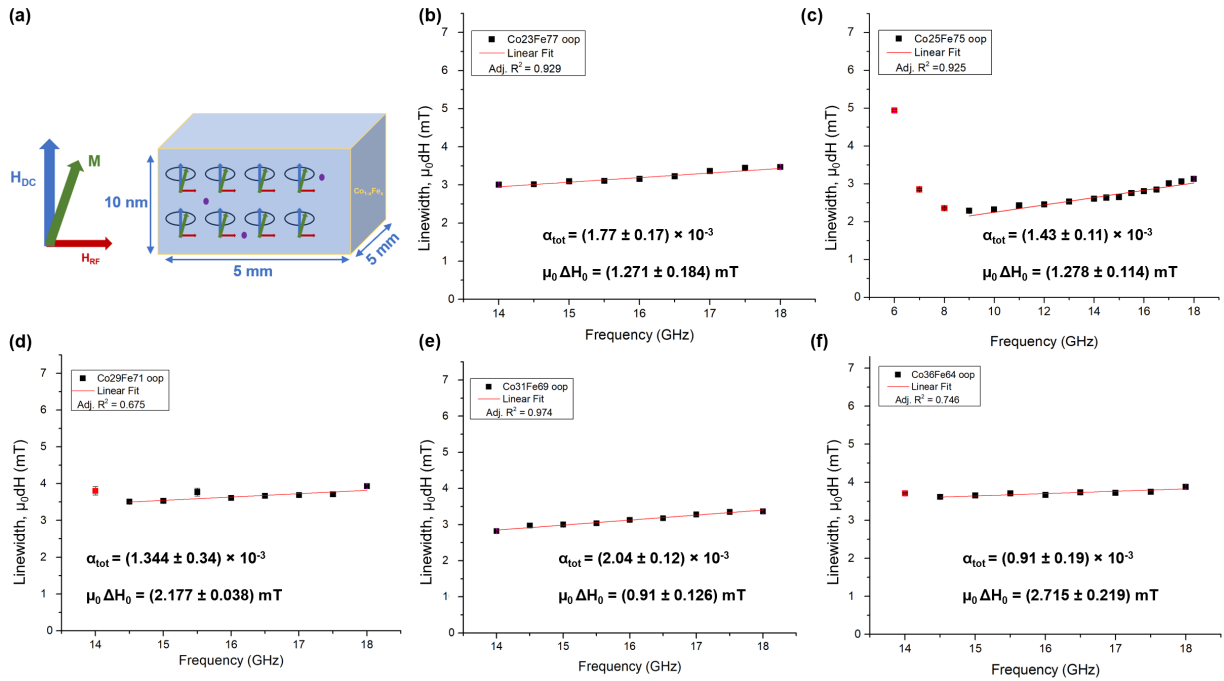

**Figure 2.** Out-of-plane linewidth measurements vs. frequency plots: (a) Illustration of out-of-plane FMR measurements and magnetization dynamics. The purple dots represent potential defects present in the film, the defect density on account of the sample thickness (10 nm) is orders of magnitude smaller than along the film plane (5 mm), b) $Co_{23}Fe_{77}$, (c) $Co_{25}Fe_{75}$, (d) $Co_{29}Fe_{71}$, (e) $Co_{31}Fe_{69}$ and (f) $Co_{36}Fe_{64}$. Note, data (open squares, red) were excluded as the external field employed at those frequencies is unable to fully align the magnetization along the hard axis (see main text).

The oop linewidths measurements shown in Figure 2 are significantly lower than those obtained in the ip geometry and for each composition the measured linewidths vs. frequency ($14 – 18$ GHz) change little. This is attributed to suppression of two magnon scattering in the oop geometry. Assuming that the density of structural or magnetic defects in the RF excitation volume ($\sim 1 \times 10^{-5}$ mm³) of the thin film is homogeneous, spin waves propagating along the film thickness (10 nm in oop FMR) experience significantly lower two magnon scattering events than when propagating along the plane of the film (5 mm in ip FMR). In a magnetic sample, the uniform FMR mode ($k = 0$, where k is the spin wavevector) is such that all spins precess in phase upon application of an external magnetic field.[13] Due to the presence of inhomogeneities in the volume of the film, additional non-uniform modes arise due to scattering and/or reflections from defects.[14] They act as centers for two-magnon scattering in which the k = 0 mode degenerates into non uniform modes that precess with a finite phase difference as compared to the uniform mode.

*Corresponding author, eemarinero@purdue.edu

As for the case of ip FMR measurements, the data shown in Figure 2 is employed to derive magnetic damping parameters in oop FMR as well (from the slope of the linewidth vs. frequency) and inhomogeneous broadening, $\mu_0 \Delta H$ (from the intercept with the y-axis of the linear fit of linewidth vs. frequency). The values of the magnetic damping parameter obtained for $Co_{25}Fe_{75}$ $(1.43 \pm 0.11) \times 10^{-3}$ compares well to the reported literature value $(1.8 \pm 0.2) \times 10^{-3}$, for the same sample composition.[7] A lower magnetic damping parameter than the published result for $Co_{25}Fe_{75}$ was measured in $Co_{36}Fe_{64}$ $(0.91 \pm 0.19) \times 10^{-3}$. The measurements also show a decreasing trend in the value of magnetic damping with increasing Co content, with the exception of $Co_{31}Fe_{69}$ sample, which exhibits the highest damping parameter, $(2.04 \pm 0.12) \times 10^{-3}$, of the samples studied.

Table 1. Magnetic properties and damping parameter measurements for in-plane and out-of-plane geometries. $\mu_0 M_s$ = saturation magnetization, $\alpha_{ip}$ = in-plane magnetic damping, $\alpha_{oop}$ = out-of-plane magnetic damping, ip $\mu_0 M_{eff}$ = in-plane effective magnetization, oop $\mu_0 M_{eff}$ = out-of-plane effective magnetization, $\mu_0 H_u^{IP}$ = in-plane uniaxial anisotropy, oop $\mu_0 H_u^{\perp}$ = out-of-plane perpendicular uniaxial anisotropy, ip $\mu_0 \Delta H_0$ = in-plane inhomogeneous broadening, oop $\mu_0 \Delta H_0$ = out-of-plane inhomogeneous broadening. Note that $M_{eff} = M_s - H_u^{\perp}$ for the calculation of $H_u^{\perp}$ in both ip and oop geometries refer to the *supplementary information*.

| Sample | $\mu_0 M_s$ (T) | $\alpha_{ip}$ $\times 10^{-3}$ | $\alpha_{oop}$ $\times 10^{-3}$ | ip $\mu_0 M_{eff}$ (T) | oop $\mu_0 M_{eff}$ (T) | $\mu_0 H_u^{IP}$ (mT) | ip $\mu_0 H_u^{\perp}$ (T) | oop $\mu_0 H_u^{\perp}$ (T) | ip $\mu_0 \Delta H_0$ (mT) | oop $\mu_0 \Delta H_0$ (mT) |
|---|---|---|---|---|---|---|---|---|---|---|
| $Co_{23}Fe_{77}$ | 1.994 ± 0.0015 | 5.61 ± 0.04 | 1.77 ± 0.17 | 1.89 ± 0.0018 | 1.939 ± 0.0018 | -1.39 ± 0.049 | 0.104 ± 0.005 | 0.055 ± 0.015 | 0.055 ± 0.052 | 1.271 ± 0.184 |
| $Co_{25}Fe_{75}$ | 2.215 ± 0.0016 | 5.35 ± 0.04 | 1.43 ± 0.11 | 1.92 ± 0.0018 | 1.966 ± 0.006 | -1.443 ± 0.048 | 0.295 ± 0.006 | 0.249 ± 0.006 | 0.042 ± 0.04 | 1.278 ± 0.114 |
| $Co_{29}Fe_{71}$ | 1.973 ± 0.0015 | 5.07 ± 0.05 | 1.344 ± 0.34 | 1.945 ± 0.0015 | 1.99 ± 0.001 | -0.0075 ± 0.038 | 0.028 ± 0.005 | - 0.017 ± 0.014 | 0.154 ± 0.041 | 2.177 ± 0.038 |
| $Co_{31}Fe_{69}$ | 1.91 ± 0.0014 | 5.53 ± 0.15 | 2.04 ± 0.12 | 1.917 ± 0.0022 | 1.961 ± 0.001 | -1.286 ± 0.059 | - 0.007 ± 0.006 | - 0.051 ± 0.009 | 0.363 ± 0.119 | 0.91 ± 0.126 |
| $Co_{36}Fe_{64}$ | 2.036 ± 0.0015 | 6.26 ± 0.09 | 0.91 ± 0.19 | 1.921 ± 0.0027 | 1.923 ± 0.0026 | -2.235 ± 0.073 | 0.115 ± 0.006 | 0.113 ± 0.004 | 0.13 ± 0.077 | 2.715 ± 0.219 |

Table 1 summarizes the magnetic and FMR measurements of this study. The saturation magnetization ($\mu_0 M_s$) value of $Co_{25}Fe_{75}$ is ~ 2.215 T. We note that this is lower than that reported in the literature: 2.4 T.[7,11] A plausible reason for the discrepancy is the accuracy in measuring the alloy thin film thickness and the determination of the interfacial "magnetic dead layer" (see *supplementary information*). Also, a lower chamber base pressure ($< 4 \times 10^{-8}$ Torr) was employed in refs 7 and 11 for sputtering $Co_{25}Fe_{75}$ thin film as compared to this work ($8.8 \times 10^{-8}$ Torr). Contaminants are well known to reduce the saturation magnetization in alloy thin films.

The results of Table 1 indicate that increasing the Co content in Co-Fe alloy, reduces the measured in-plane magnetic damping parameter from $Co_{23}Fe_{77}$ to $Co_{29}Fe_{71}$ and then increases from $Co_{31}Fe_{69}$ to a maximum in $Co_{36}Fe_{64}$. Likewise, as the Co content in $Co_{1-x}Fe_x$ series increases, the magnetic damping constant measured in oop geometry decreases, with the notable exception of $Co_{31}Fe_{69}$ for which the largest damping parameter in this geometry is observed. Measurements of the out-of-plane magnetic anisotropy, yield the highest values of $H_u^\perp = 0.295$ T (ip FMR) and 0.249 T (oop FMR) for $Co_{25}Fe_{75}$. This is likely due to contributions from interface magnetic anisotropy. This perpendicular anisotropy enables the external magnetic field to more effectively orient the magnetization along the hard axis. As a result, the RF power absorption vs. $\mathbf{H_{DC}}$ is narrower, reducing the linewidth (dH) and the slope in dH vs. frequency. This yields a lower total magnetic damping constant for this composition. In the table $\boldsymbol{\mu_0 H_u^{IP}}$ is extracted from the FMR measurements in in-plane geometry.

*Corresponding author, eemarinero@purdue.edu

Conversely, for $Co_{31}Fe_{69}$, $H_u^\perp = $ - 0.007 T (ip FMR) and - 0.051 T (oop FMR), the negative and very small values indicate absence of out-of-plane anisotropy. Thus, this sample, is harder to fully magnetize along the hard axis than the counterpart films exhibiting larger and positive values. The saturation process requires overcoming the demagnetization of the bulk of the sample and any in-plane magnetocrystalline anisotropy present in the film. This results in a wider microwave power absorption in frequency range and a wider linewidth (dH) leading to a higher value of the magnetic damping constant. The results for $Co_{36}Fe_{64}$ are somewhat surprising: this sample exhibits the lowest magnetic damping parameter of the series. As the magnitude of $H_u^\perp = 0.115$ T (ip FMR) for this composition is lower than that of $Co_{25}Fe_{75}$ ($H_u^\perp = 0.295$ T), it cannot be argued that the interface magnetic anisotropy for this sample enables a more facile hard axis orientation. Other contributions such as strain-lattice relaxation are considered in this work based on HAADF-STEM analysis.

The inhomogeneous linewidth broadening measurements reported in Table 1 for oop geometry FMR measurements are significantly larger than those derived from ip geometry FMR configuration. Whereas linewidth broadening due to two magnon scattering originating from defects in the ferromagnetic layer are drastically suppressed due to the layer thickness in oop FMR measurements, local variations of magnetic properties across the thickness of the magnetic layer play a larger role.[15] At the Cu seed interface, interdiffusion and chemical interactions results in a "magnetic dead layer" of ~ 0.474 nm in the nominal 10 nm thick films as shown in Figure S2 of *supplementary information*. In addition, Cu interdiffusion into the magnetic layer was revealed by the STEM analysis conducted in this work. Such atomic level inhomogeneities (magnetic and compositional) and interfacial contributions are responsible for incrementing the inhomogeneous linewidth. As observed in Table 1, there is no obvious correlation between the magnetic damping parameters measured and the inhomogeneous linewidths. The inhomogeneous linewidths between $Co_{36}Fe_{64}$ and $Co_{29}Fe_{71}$ are comparable (20 %) and significantly different magnetic damping parameters (48 %) in oop FMR. One could argue that the inhomogeneous linewidth measured is a consequence of the limited FMR frequency range of our study. However, the impact of measurement frequency on inhomogeneous linewidth broadening can be expected to be important at low frequencies when the applied field is insufficient to align the magnetization along the hard axis. As indicated earlier, above 14 GHz the fields applied exceed the saturation magnetization of the samples. Another factor that could contribute to the inhomogeneous linewidth in the polycrystalline alloys is a spread of crystalline growth orientations which could result in variations of local magnetic properties leading to the broadening of the linewidth. FMR measurements at higher frequencies would be desirable to improve the accuracy for the extracted values of magnetic damping and inhomogeneous linewidth. However, we do not have this capability and will pursue such measurements in the future through collaborations. We note that the work of Lee et al.[10] on the epitaxial growth of $Co_{25}Fe_{75}$ on MgO and $MgAl_2O_4$ substrates reports magnetic damping and inhomogeneous linewidths values of $7.1 \times 10^{-4}$ and 9.0 Oe vs. $1.0 \times 10^{-3}$ and 1.9 Oe respectively. It is interesting to note that in this study also, the values of magnetic damping and linewidth are not correlated.

*Corresponding author, eemarinero@purdue.edu

The effective magnetization values measured in oop and ip FMR configurations shown in Table 1 differ from 0.2 % to 2.6 %. Several factors could contribute to the discrepancy this includes sample alignment with respect to the direction of the applied field. Another possibility is the absence of higher frequency measurements in the oop-geometry measurements on account of our limited FMR capabilities.

## HAADF STEM Microstructural analysis

To provide experimental information on the correlation between the thin film microstructure and the measured magnetic properties, cross sectional STEM analysis was performed. The objective of the study was to measure lattice parameters, crystallographic orientations and elemental analysis with atomic resolution of the $Co_{36}Fe_{64}$ thin films and of the Cu buffer layers.

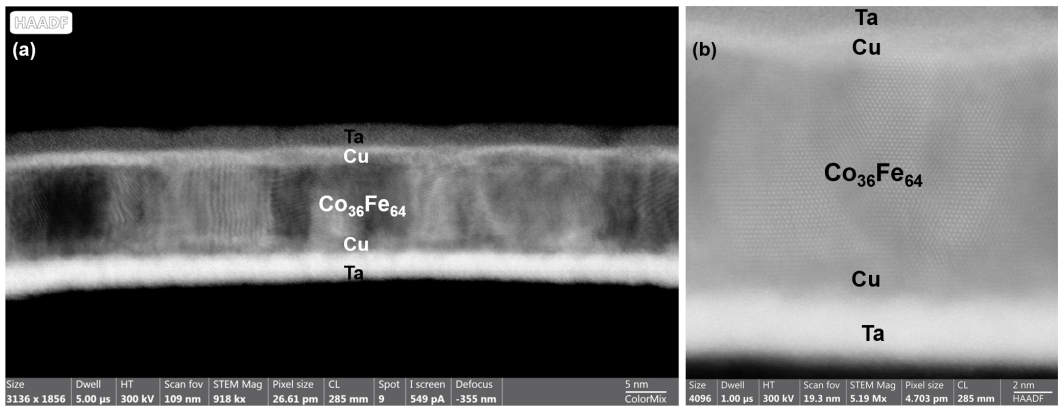

**Figure 3.** STEM images of thin film stacks with $Co_{36}Fe_{64}$ magnetic layers: (a) cross section of the thin film stack, (b) higher resolution image, lattice fringes and grain morphology are readily observable.

Figure 3 (a) was obtained with a 918 kx magnification with total scan field of view of 109 nm. The polycrystalline growth of the $Co_{36}Fe_{64}$ layer is readily observed and the sample consists of multiple grains separated by grain boundaries. Figure 3 (b) is a magnified version of Figure 3 (a) and the magnification employed was 5.19 Mx.

*Corresponding author, eemarinero@purdue.edu

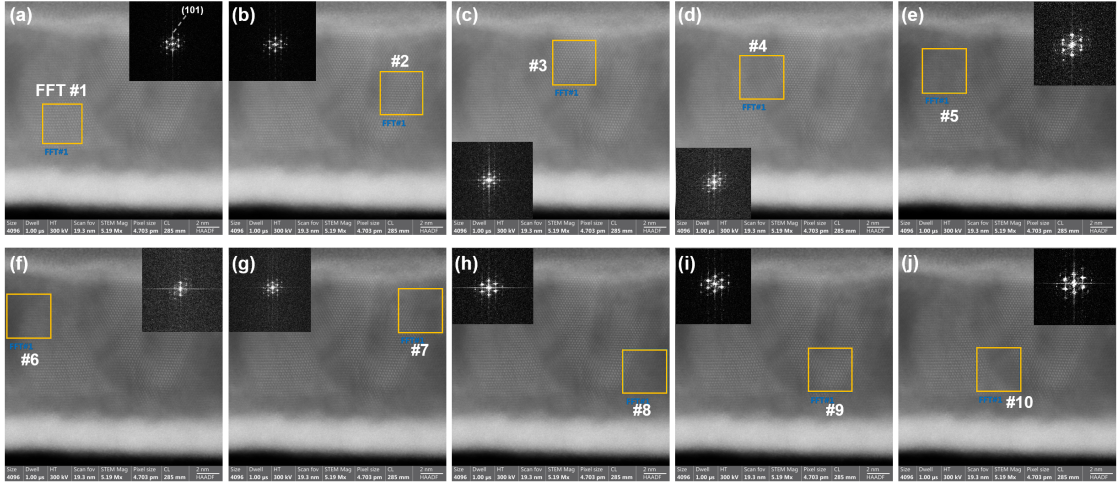

**Figure 4.** Diffraction analysis of Co$_{16}$Fe$_{64}$ layer. Insets provide the HAADF diffraction patterns at the particular location analyzed (yellow boxes) in Figures (a) to (j). The dashed line in the diffraction pattern in the inset of (a) indicates the direction of the (101) growth plane in Co$_{36}$Fe$_{64}$ as discussed in the main text.

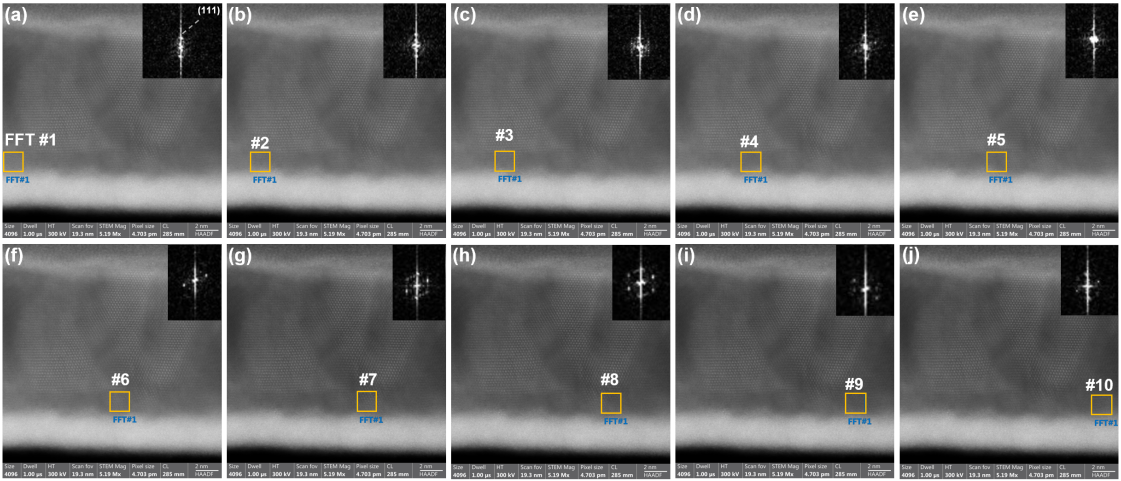

**Figure 5.** Diffraction analysis of Cu buffer layer. Insets provide the HAADF diffraction patterns at the particular location analyzed (yellow boxes) in Figures (a) to (j). The dashed line in the diffraction pattern in the inset of (a) represents the direction of the growth plane (111) for Cu. See main text for details on this determination.

Diffraction analysis was conducted to calculate the inter planar spacing (d-spacing) of the observed diffraction patterns. Velox software was employed for diffraction analysis of the captured STEM images by computing Fast Fourier Transforms (FFT) of the selected sections (yellow boxes) in Figures 4 and 5. Several grain areas were measured to obtain an aggregate characterization of the diffraction patterns. The dimensions of the analyzed areas were identical within each layer to ensure consistency in the analysis. Figure 4 presents locations of different grain areas analyzed in the Co$_{36}$Fe$_{64}$ layer along with their respective diffraction patterns. From this study, it was deduced that, the plane along the primary growth direction of Co$_{36}$Fe$_{64}$ is (101) and is marked by a dashed grey line in the inset of Figure 4 (a). Similarly, Figure 5 provides the locations of grain areas in Cu buffer layer that were analyzed and diffraction patterns are also presented in the inset of each image. It was found that the plane along the primary growth direction of Cu buffer layer is (111) and is identified in Figure 5 (a) by a dashed grey line in the inset of the figure. The description on the identification of the growth planes for both Cu and $Co_{36}Fe_{64}$ is presented next.

*Corresponding author, eemarinero@purdue.edu

**Table 2.** Summary of interplanar spacing values in the growth directions (perpendicular to the substrate plane) for $Co_{36}Fe_{64}$ (101) and Cu (111) buffer layer.

| | $Co_{36}Fe_{64}$ | | Cu |
|---|---|---|---|
| FFT # | d (101) (nm) | FFT # | d (111) (nm) |
| 1 | 0.237699 | 1 | 0.218364 |
| 2 | 0.216193 | 2 | 0.225785 |
| 3 | 0.21645 | 3 | 0.214385 |
| 4 | 0.216146 | 4 | 0.213174 |
| 5 | 0.215564 | 5 | 0.214064 |
| 6 | 0.221754 | 6 | 0.214938 |
| 7 | 0.227402 | 7 | 0.214385 |
| 8 | 0.243605 | 8 | 0.213881 |
| 9 | 0.232126 | 9 | 0.213767 |
| 10 | 0.243309 | 10 | 0.229489 |

Using the measured parameters tabulated in Table 2, the d-spacing and lattice parameters for both Cu and $Co_{36}Fe_{64}$ were calculated using Equation 5[16]:

$$d_{hkl} = \frac{a}{\sqrt{h^2+k^2+l^2}} \qquad \text{(Equation 5)}$$

where $d_{hkl}$ is d-spacing and $h, k, l$ corresponds to the planes Miller indices and $a$ is the lattice parameter. To identify the growth planes, average values of d-spacing were taken from Table 2 for both Cu and $Co_{36}Fe_{64}$ and then the lattice parameters of Cu[17] and $Co_{38}Fe_{62}$[16] were taken from the literature. Since no lattice parameters were found for $Co_{36}Fe_{64}$ so the lattice parameter of the next nearest composition $Co_{38}Fe_{62}$ was considered. Then using the Equation 5, planes with possible matching were identified and the planes that closely matched the values of d-spacing obtained for Cu and $Co_{36}Fe_{64}$ were identified as (111) and (101) respectively. For precise identification of the growth plane for $Co_{36}Fe_{64}$, standard TEM diffraction patterns for the bcc crystal structures were used from a database.[18]

For Cu, the d-spacing for (111) plane derived from FFT analysis of regions #3 to #9 was calculated to be 0.214085 nm. Similarly, for $Co_{36}Fe_{64}$ layer, d-spacing for (101) plane derived from FFT analysis of regions #2 to 5 was calculated to be 0.216088 nm. Other FFT areas analyzed in Cu and $Co_{36}Fe_{64}$ were not considered on account of excessive intermixing. Evidence of intermixing is presented later in the elemental analysis discussion. These d-spacing values for both Cu and $Co_{36}Fe_{64}$ were employed together with Equation 5 to calculate their lattice constants. The lattice constant for Cu was determined to be 0.37081 nm and for $Co_{36}Fe_{64}$ was 0.30559 nm.

Since the crystallographic planes in the growth direction and those parallel to the substrate are perpendicular to each other, it is possible to identify the parallel planes that provide lattice matching between the Cu buffer layer and $Co_{36}Fe_{64}$.

*Corresponding author, eemarinero@purdue.edu

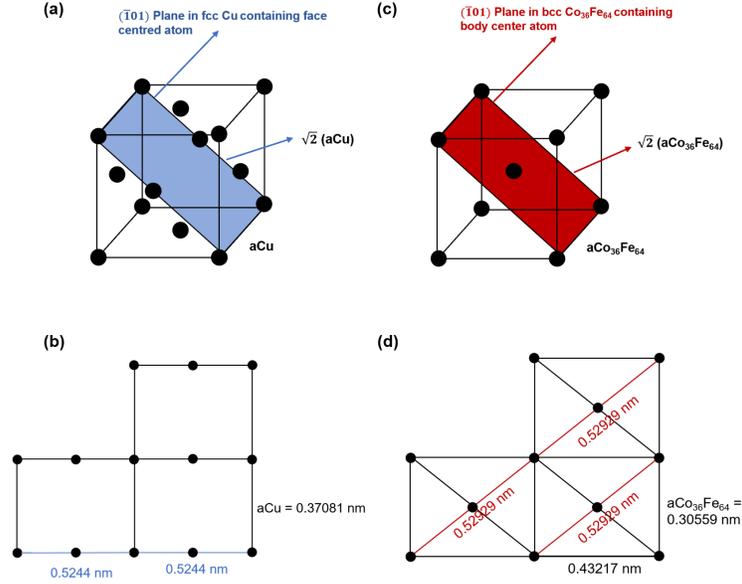

**Figure 6.** Illustration of the main crystallographic planes of $Co_{36}Fe_{64}$ grown on Cu buffer layers: (a) the $(\bar{1}01)$ plane in fcc Cu with corner and face center atoms, (b) 2D projection of the $(\bar{1}01)$ planes of adjacent fcc Cu unit cells; (c) the $(\bar{1}01)$ plane in bcc $Co_{36}Fe_{64}$ containing corner and body center atoms; (d) 2D projection of the $(\bar{1}01)$ planes of adjacent bcc $Co_{36}Fe_{64}$ unit cells.

For both, Cu and $Co_{36}Fe_{64}$, the $(\bar{1}01)$ planes are perpendicular to their respective growth direction planes (111) and (101) respectively. The $(\bar{1}01)$ plane for bcc $Co_{36}Fe_{64}$ and $(\bar{1}01)$ for fcc Cu are identified in their respective unit cells in Figure 6 (c) and 6 (a) respectively. The Cu $(\bar{1}01)$ plane includes face center and corner atoms, whereas in bcc $Co_{36}Fe_{64}$ it includes the body center atom and corner atoms. 2D projections of these planes for fcc Cu and bcc $Co_{36}Fe_{64}$ are given in Figures 6 (b) and 6 (d) respectively. The diagonal of the $(\bar{1}01)$ plane in bcc $Co_{36}Fe_{64}$ has a length of 0.52929 nm and the longer edge of the of the $(\bar{1}01)$ plane in fcc Cu has the length of 0.5244 nm. From these considerations, the lattice mismatch between $Co_{36}Fe_{64}$ and Cu buffer layer is estimated to be 0.9325 % and is calculated from the expression below:

$$\frac{|diagonal\ of\ Co36Fe64\ (\bar{1}01) - edge\ of\ Cu\ (\bar{1}01)\ plane|}{edge\ of\ Cu\ (\bar{1}01)\ plane} \times 100 \quad \text{(Equation 6)}$$

Kharmouche and Melloul[16] have measured lattice parameters of thermally evaporated $Co_{1-x}Fe_x$ thin films on monocrystalline Si <111> substrate. The Co composition range studied was from 38 % to 65 %. Their measurements are shown in Figure 7. A monotonic decrease in lattice constant is observed as the Co content is incremented. The linearity of the variation in lattice constant enables us to estimate by extrapolation the lattice constants that would be observed in their experiment for the compositional range studied in this work. The HAADF diffraction analysis in our work of $Co_{36}Fe_{64}$ grown on Cu yielded a lattice constant of 0.30559 nm. The value derived from extrapolation of the measurements in reference 16 for this composition is 0.28513 nm. The 7 % difference is likely due to the difference in thin film growth methods (thermal evaporation vs. sputtering) and our films are grown on Cu underlayers to induce the crystallographic texture in the CoFe alloy. Both sets of CoFe alloys prepared in ref 16 and in our work are bcc. Therefore, we can assume that the lattice constants of our compositional range would display the same linear behavior, namely a reduction of the lattice constant with increasing Co content. This enables us to derive a graphic approximation of the lattice constants for all CoFe compositions of our study by drawing a line through our measured data point with the same slope as derived in the measurements of Kharmouche and Melloul, as shown in Figure 7.

*Corresponding author, eemarinero@purdue.edu

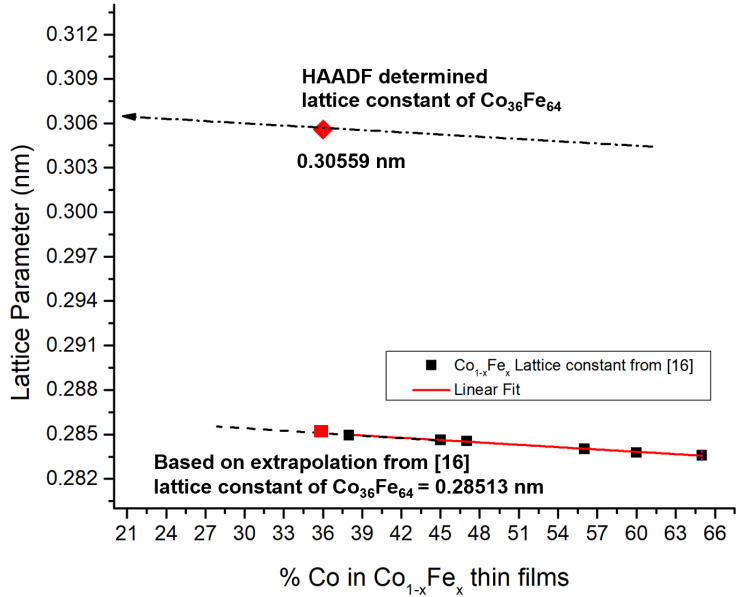

**Figure 7.** Lattice parameters of $Co_{1-x}Fe_x$ thin films as a function of % Co grown by thermal evaporation on Si. Black boxes are the experimental values reported in reference 16, the red box is the extrapolated value for a $Co_{36}Fe_{64}$ sample grown under their experimental conditions. The red diamond is the value of lattice constant of $Co_{36}Fe_{64}$ grown on Cu by sputter deposition in our experiment and measured by HAADF analysis. A line with the same slope as the reference data drawn through our experimental point enables an estimate of the lattice constants for the composition series studied in this work.

The values of the lattice parameters for the $Co_{1-x}Fe_x$ compositions of this study derived from the extrapolation are tabulated in Table 3. The diagonal lengths of their respective $(\bar{1}01)$ planes were derived using the same procedure used for the experimental result for $Co_{36}Fe_{64}$. The lattice parameter for fcc Cu is assumed to be constant and since all samples were grown under identical conditions. We use in our evaluation the degree of lattice matching from the experimental result derived from the HAADF analysis. Lattice mismatch values were calculated using Equation 6 and are provided in the 4th column of Table 3.

It is evident from Table 3 that upon increasing % Co from 23 % to 36 %, the % lattice mismatch reduces from 1.18 % in $Co_{23}Fe_{77}$ to 0.93 % for the case of $Co_{36}Fe_{64}$. Overall, this is indicative of excellent lattice matching between $Co_{1-x}Fe_x$ and Cu buffer layer.

*Corresponding author, eemarinero@purdue.edu

**Table 3.** Interpolated values of $Co_{1-x}Fe_x$ lattice constants (column 2); values of the diagonal lengths of the $(\bar{1}01)$ plane of bcc $Co_{1-x}Fe_x$ (column 3) and the % lattice mismatch between the diagonal of $(\bar{1}01)$ plane of $Co_{1-x}Fe_x$ and edge of Cu $(\bar{1}01)$ plane vs. Co content (column 4).

| Sample | Interpolated Lattice constant (nm) | Diagonal of $Co_{1-x}Fe_x$ $(\bar{1}01)$ plane to match the edge of Cu $(\bar{1}01)$ plane (nm) | % Lattice mismatch with Cu layer |
|---|---|---|---|
| $Co_{23}Fe_{77}$ | 0.30636 | 0.53063 | 1.188 |
| $Co_{25}Fe_{75}$ | 0.30625 | 0.53044 | 1.1518 |
| $Co_{29}Fe_{71}$ | 0.30608 | 0.53015 | 1.0965 |
| $Co_{31}Fe_{69}$ | 0.30597 | 0.52995 | 1.0583 |
| $Co_{36}Fe_{64}$ | 0.30559 [HAADF] | 0.52929 | 0.9325 |

The HAADF diffraction analysis revealed lattice expansion of $Co_{36}Fe_{64}$ due to Cu interdiffusion. To further investigate interdiffusion, elemental analysis was conducted using Velox software wherein elemental maps were generated at different locations of the thin film stack. Several locations of the same HAADF STEM image were investigated by placing vertical narrow rectangles at different positions of the image. The locations of the areas probed in this analysis included some areas with grain boundaries and some without them. Figures 8 (a) to (p) present the color-coded elemental maps at various locations (vertical narrow rectangular boxes) of the thin film stack. The analyzed areas were identical to ensure consistency in analysis. Figure 9 presents normalized line averaged profiles for Fe, Co, Cu and Ta from the analysis of Figure 8. Figures 9 (a) to (p) correspond to the line averaged profiles generated at each position from position #1 to 16 respectively. The amplitude of the line averaged profiles corresponds to the percent atomic fraction of each element present in the narrow vertical region being probed. The right-hand side of each plot of Figure 9 present the line profiles near the $Si\text{-}SiO_2$ substrate, and those on the left-hand side correspond to top layer elements. This convention is shown in Figure 9 (a). Line profiles represent the atomic fraction (%) of each element as labelled on y-axis of the plots. For the purpose of quantification of their percent atomic fractions at a particular location, an average is taken from several data points of each line profile.

*Corresponding author, eemarinero@purdue.edu

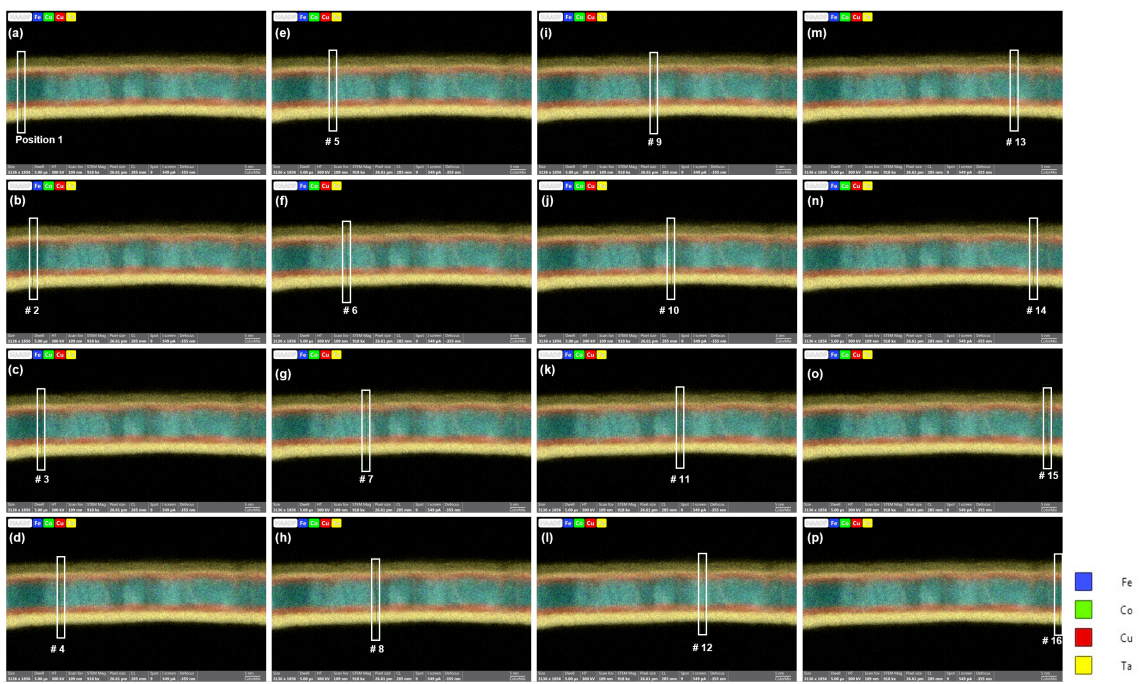

**Figure 8.** Color-coded STEM images of Co, Fe, Ta and Cu. Figures from (a) to (p) show positions of different cross-sectional areas probed from position #1 to position # 16 respectively to perform elemental analysis. A color legend on the right presents respective elements in the line plots.

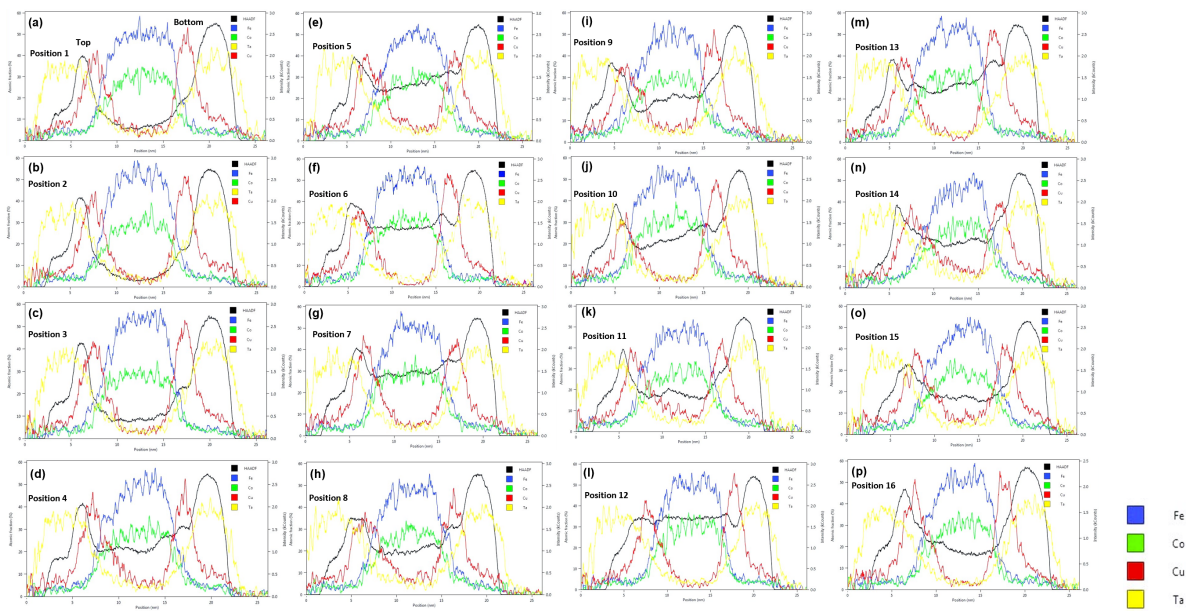

**Figure 9.** Normalized line averaged profiles of the respective areas analyzed in Figure 8. Position numbers in the plots from (a) to (p) represent their respective box positions from position 1 to 16 respectively in Figure 8. In the plots, peaks to the right of the plot layers are close to the substrate and peaks towards the left are towards the top of the stacks and away from the substrate as shown in Figure 9 (a). A color legend on the right presents respective elements in the line plots.

*Corresponding author, eemarinero@purdue.edu

For simplicity, only the Fe and Cu signals are considered in the analysis of Cu interdiffusion. The Co profile in the magnetic layer follows a similar, but different in amplitude, distribution shape as Fe across the thickness of the $Co_{36}Fe_{64}$ film. A plot summarizing the extent of Cu interdiffusion is presented in Figure 10. To determine the extent of Cu interdiffusion, the ratio of the atomic fraction of Cu in $Co_{36}Fe_{64}$ to the fraction of Fe in $Co_{36}Fe_{64}$, labelled as Cu-CoFe/Fe on the y-axis, is plotted vs the analysis location.

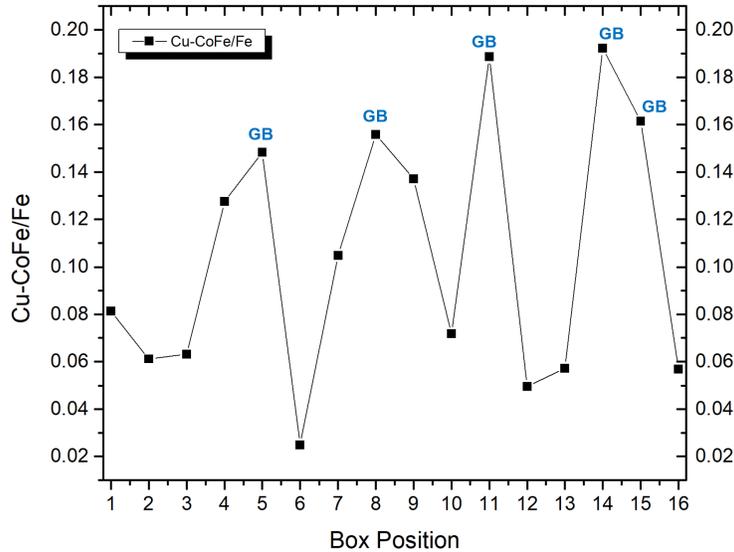

**Figure 10.** Cu interdiffusion analysis: The ratio of the atomic fractions of Cu in $Co_{36}Fe_{64}$ to Fe in the $Co_{36}Fe_{64}$ layer is plotted vs the analyzed areas in Figure 8. Note that GB in the plot refers to grain boundary.

The plot in Figure 10 clearly reveals that the Cu/Fe ratio is highest at grain boundaries (locations 5, 8, 11, 14 and 15) and a minimum in the center of the CoFe grains (locations 6, 12, 13 and 16).

Moreover, Cu interdiffusion can also be visualized by observing Cu content (red line) in line profile plots in Co-Fe region (mid region) of Figure 9. In those regions of plots, Cu content varies from almost negligible (positions 6, 12, 13, 16) to as high as ~ 10 % (positions 5, 8, 11, 14, 15) in some areas. Hence, it is clear that Cu interdiffusion takes place into the $Co_{36}Fe_{64}$ layer via the grain boundaries and its extent is highest at the grain boundaries as compared to other areas in grains.

Energy-dispersive X-ray Spectroscopy (EDS) analysis is presented in Figure 11 for each layer of the multilayered thin film stack of the $Co_{36}Fe_{64}$ sample.

*Corresponding author, eemarinero@purdue.edu

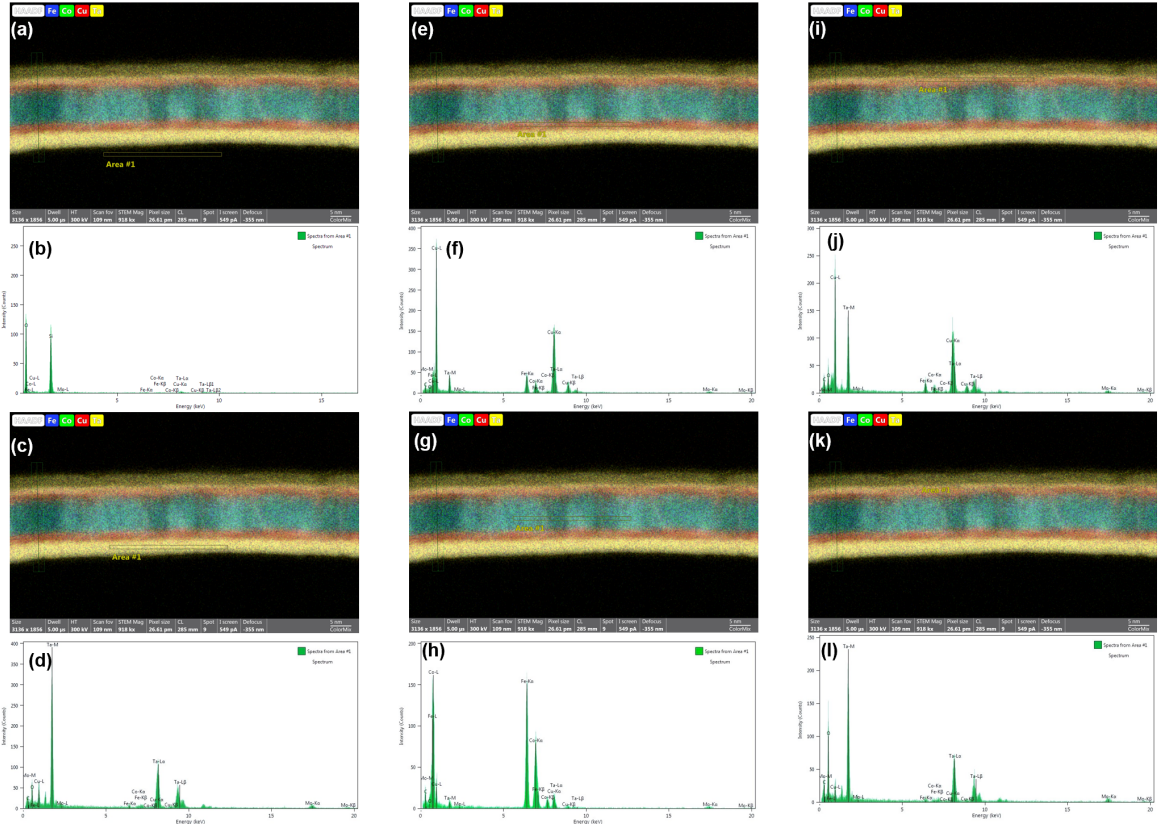

**Figure 11.** Elemental maps across the sample stack of the $Co_{36}Fe_{64}$ sample: (a) analysis area in Si-SiO₂ substrate; (b) EDS spectrum of Si-SiO₂ substrate; (c) analysis area in Ta seed layer; (d) EDS spectrum of Ta seed layer; (e) analysis area in Cu bottom layer; (f) EDS spectrum of Cu bottom layer; (g) analysis area in $Co_{36}Fe_{64}$ layer; (h) EDS spectrum of $Co_{36}Fe_{64}$ layer; (i) analysis area in Cu top layer; (j) EDS spectrum of top Cu layer; (k) analysis area in Ta cap layer and (l) EDS spectrum of Ta cap layer.

To further study Cu interdiffusion, EDS analysis was conducted of each stack layer and the Si-SiO₂ substrate in the multilayered stack using Velox software. A narrow horizontal box was placed in each layer and EDS analysis was done to determine the elemental composition of each layer. Of note, the dimensions of each box were kept identical to ensure consistency in analysis. Results from the EDS analysis are presented in Figure 11. Location of each horizontal box is shown in Figure 11 (a), (c), (e), (g), (i) and (k) pertaining to the Si-SiO₂ substrate, the Ta seed layer, the Cu buffer layer, the $Co_{36}Fe_{64}$ layer, the Cu top layer and the Ta cap layer respectively. Their respective EDS spectra are given in parts (b), (d), (f), (h), (j) and (l) respectively.

After analyzing the EDS spectrum in part (l) it is clear that Ta cap layer is mostly oxidized which is expected since it is the top layer in the thin film stack. Moreover, Cu is observed in $Co_{36}Fe_{64}$ layer as is evident from the EDS spectrum in part (h) as Cu-Kα and Cu-Kβ lines are observed.

Hence, from the STEM analysis of $Co_{36}Fe_{64}$ composition it is evident that Cu interdiffusion takes place primarily through the grain boundaries. There exists strong magnetic exchange interaction within the grains of ferromagnetic thin film which is critical for efficient magnon propagation. Due to Cu interdiffusion into the Co-Fe film, this exchange interaction is disrupted as Cu atoms in the ferromagnetic thin film act as non-magnetic defect centers. This phenomenon eventually results in incrementing the magnetic damping parameter of the ferromagnetic $Co_{1-x}Fe_x$ thin films.

*Corresponding author, eemarinero@purdue.edu

EDS Analysis of the entire series studied would be needed to find out whether the changes in magnetic damping directly correlate to differences in Cu interdiffusion in the different samples.

In addition, to study the effect of reducing outgassed contaminants in the sputter deposition chamber on magnetic damping, prior to the deposition of the thin film stacks, Ta gettering was employed to lower the base pressure of the sputter chamber from $9 \times 10^{-8}$ Torr down to $5 \times 10^{-8}$ Torr. Ta gettering is a known as an efficient procedure to remove contaminants such as oxygen and water vapor from the vacuum chamber.[19,20] The lower base pressure (reduced adventitious contaminants) reduced the in-plane magnetic damping parameters by around 18 % indicative that residual contaminants during sputter deposition are deleterious and that lower chamber base pressures could potentially reduce the magnetic damping even further.

**Conclusion**

We have investigated the compositional dependence of magnetic damping in $Co_{1-x}Fe_x$ polycrystalline thin films spanning the Co content in the vicinity of the composition $Co_{25}Fe_{75}$, previously reported to exhibit ultralow magnetic damping[7]. FMR measurements using the out-of-plane geometry to reduce two magnon scattering contributions, yielded a value of $\sim 0.91 \times 10^{-3}$ in 10 nm thick $Co_{36}Fe_{64}$ thin films. The ferromagnetic layers were grown by co-sputtering at ambient temperature and intercalated within Cu(3 nm)/Ta(3 nm) bilayers, the thin film stacks were deposited on thermally oxidized Si wafers. We observed a decreasing trend in the magnetic damping parameter with increasing % Co content in the binary alloys except for $Co_{31}Fe_{69}$ which exhibited the highest magnetic damping parameter ($\sim 2.04 \times 10^{-3}$) of the series. Magnetic measurements revealed that this alloy composition has no measurable interface perpendicular magnetic anisotropy (IPMA) whereas the measurements in $Co_{36}Fe_{64}$ determined significant IPMA. The presence of IPMA facilitates the magnetization alignment in the hard axis direction. It is noted that the highest IPMA was measured in the $Co_{25}Fe_{75}$ but its magnetic damping parameter was not the lowest ($\sim 1.43 \times 10^{-3}$). The HAADF-STEM analysis provided additional information to account for the Co compositional dependence of magnetic damping reported in this study. The diffraction analysis identified the crystallographic planes of bcc $Co_{36}Fe_{64}$ and fcc Cu in both the growth direction and along the substrate plane. This enabled us to determine the lattice mismatch between the ferromagnet and the Cu seed layer. By extrapolation and utilizing published results of lattice parameters for thermally evaporated $Co_{1-x}Fe_x$ alloys[16], the lattice mismatch as a function of Co content was estimated and found to decrease as the Co content of the $Co_{1-x}Fe_x$ is incremented. Elemental analysis of $Co_{36}Fe_{64}$ with atomic resolution from STEM cross sectional studies revealed interdiffusion of Cu into the magnetic layer, which was found to be 7x higher at the grain boundaries as compared the bulk of the polycrystalline grains in the ferromagnetic film.

The static and dynamic magnetic measurements together with the HAADF-STEM structural analysis enables us to suggest the following plausible mechanism for the measured Co compositional dependence observed in our study. Ultralow magnetic damping in $Co_{1-x}Fe_x$ alloys in the vicinity of the Co = 25 % composition as discussed by Schoen et al.[7] arises from a sharp minimum of the density of states at the Fermi level. However, magnetic damping of spin waves in nanoscale thin film $Co_{1-x}Fe_x$ alloys is strongly influenced by their structural, compositional and interface properties. The large differences in magnetic damping parameters measured along the thin film plane (5 mm) vs the thickness (10 nm) evidence the negative influence of magnon scattering on account of the large difference in structural defects and magnetic fluctuations in the probed FMR volume. The magnetic damping measured in both ip and oop configurations decreases as the Co content is incremented (23 % - 29 %). The thin film strain decreases in this composition range, thus a decrease in Cu interdiffusion can be expected, whereas the presence of IPMA favors the magnetization alignment in the hard axis for oop-geometry FMR measurements. IPMA is virtually absent for the case of the Co = 29 % film which favors in-plane magnetization alignment. Table 1 also shows that the $\mu_0 H_u^{IP}$ = in-plane uniaxial anisotropy (IUA) also is reduced in this composition range, this also favors the magnetization orientation in plane, resulting in a lower magnetic damping parameter. This reveals the counteracting effects from interdiffusion, IPMA and IUA on magnetic damping measurements in this composition range. In the case of Co = 31 %, an increase in magnetic damping is observed in both measurement geometries, this sample exhibits essentially no IPMA and IUA value is larger than for Co = 29 %. These magnetic properties negatively impact magnetic damping and the influence of interdiffusion appears to be less dominant. Finally, the Co = 36 % sample exhibits both the largest and smallest magnetic damping parameter measured in both geometries. IUA is the largest of the series and the value of IPMA is the second largest of the series. This is expected to influence magnetic damping in ip and oop directions. Cu interdiffusion is likely to be the least of the series and a combination of large IPMA and low thin film strain are likely responsible for the low magnetic damping parameter measured for this sample in the oop-geometry FMR measurement.

*Corresponding author, eemarinero@purdue.edu

The key role played by impurities in the magnetic layer were confirmed in our work in which we obtained a reduction of the in-plane magnetic damping parameter as high as 18 % for $Co_{36}Fe_{64}$ composition when the samples were grown under lower sputter chamber base pressure conditions. With respect to interdiffusion, we note that the epitaxial growth of $Co_{25}Fe_{75}$ on MgO by Lee et al.[10] yielded an ultralow magnetic damping parameter of $7.1 \times 10^{-4}$. Although no cross sectional elemental compositional studies are reported, we suggest that the strong ionic bonding in MgO prevents interdiffusion of either Mg or Oxygen into the $Co_{25}Fe_{75}$ film. Metallic bonding in Cu is weaker and therefore in our thin films, Cu interdiffusion into the magnetic layer is thermodynamically favored.

This work identified potential avenues to further reduce magnetic damping in metallic ferromagnetic $Co_{1-x}Fe_x$ alloy thin films with Co content in the vicinity of Co = 25 %. The results presented are suggestive that the growth of these thin films in ultra-clean growth environments, controlling the alloy composition, impurities, thin film strain and interdiffusion from seed layers required for crystallographic control would enable possibly lower magnetic damping parameters than those reported to-date. Furthermore, for the practical implementation of magnonic devices based on these ferromagnetic materials, it highlights the importance of controlling their compositional and structural properties.

*Corresponding author, eemarinero@purdue.edu

## Supplementary Material

See the supplementary information for additional information on magnetometry measurements, dead layer determination, analysis of raw FMR data for the extraction of the magnetic damping parameter. Analysis of the Kittel equation fit to extract the effective magnetization is also provided.

## Acknowledgments

The authors would like to thank the School of Materials Engineering at Purdue University for supporting this work and also Dr. Neil Dilley for his help with the magnetic characterizations employed in this work.

*Author declarations*

**Conflict of Interest**: The authors declare no conflict of interest.

## Author Contributions

**Samanvaya S. Gaur**

Conducted all experimental work including thin film growth, magnetic, FMR measurements and data analysis, was also involved in the writing and editing of the manuscript.

**Rosa Diaz**

Conducted cross section STEM analysis and helped in its data analysis.

**Ernesto E. Marinero**

Conceived the experiment, supervised all aspects of the experimental work, provided the framework to understand the results and writing - reviewing and editing the manuscript.

The authors contributed equally to the writing of the manuscript. All authors have read and agreed to the published version of the manuscript.

**Funding:** This research was funded by the Office of Naval Research grant number [N00014-21-1-2562].

## Data availability

The data that supports the findings of this study are available from the corresponding author upon reasonable request.

*Corresponding author, eemarinero@purdue.edu

*Corresponding author, eemarinero@purdue.edu

*Corresponding author, eemarinero@purdue.edu